\documentclass[twocolumn]{openjournal}

\usepackage[utf8]{inputenc}
\usepackage[english]{babel}
\usepackage{hyperref}
\hypersetup{
    colorlinks=true,
    linkcolor=black,
    filecolor=blue,      
    urlcolor=blue,
    citecolor=blue,
}
\usepackage{color,colortbl}
\DeclareGraphicsExtensions{.bmp,.png,.jpg,.pdf}
\usepackage{verbatim}
\usepackage[normalem]{ulem}
\usepackage{orcidlink}

\newcommand{\orcidauthor}[3]{\author{\href{http://orcid.org/#1}{#2$^{#3}$}}}
\begin{document}

\title{Assessing Your Observatory's Impact:\\Best Practices in Establishing and Maintaining Observatory Bibliographies}

\orcidauthor{0000-0003-3073-0605}{Raffaele D’Abrusco}{1}
\orcidauthor{0000-0003-2975-8301}{Monique Gomez}{2}
\orcidauthor{0000-0001-6830-0702}{Uta Grothkopf}{3}
\orcidauthor{0000-0002-5858-2002}{Sharon Hunt}{4}
\orcidauthor{0000-0003-1194-9265}{Ruth Kneale}{5}
\orcidauthor{0009-0006-7491-9396}{Mika Konuma}{6}
\orcidauthor{0000-0002-8523-015X}{Jenny Novacescu}{7,*}
\orcidauthor{0000-0001-6381-515X}{Luisa Rebull}{8}
\orcidauthor{0000-0003-1468-5357}{Elena Scire}{9}
\orcidauthor{0000-0003-2622-3352}{Erin Scott}{1}
\orcidauthor{0000-0003-4058-5202}{Richard Shaw}{7}
\orcidauthor{0000-0001-6870-2365}{Donna Thompson}{10}
\orcidauthor{0009-0007-7866-8800}{Lance Utley}{11}
\orcidauthor{0009-0003-8345-463X}{Christopher Wilkinson}{7}
\orcidauthor{0000-0001-7354-6221}{Sherry Winkelman}{1}

\affiliation{$^{1}$Chandra Data Archive, Chandra X-ray Center (CXC) / Center for Astrophysics, Harvard \& Smithsonian, 60 Garden Street, Cambridge, MA 02138, USA}

\affiliation{$^{2}$Instituto de Astrofísica de Canarias (IAC), Vía Láctea, s/n E-38205, La Laguna - Tenerife, Spain}

\affiliation{$^{3}$European Southern Observatory (ESO), Karl-Schwarzschild-Straße 2, 85748 Garching bei München, Germany}

\affiliation{$^{4}$NSF NOIRLab, 950 N Cherry Ave, Tucson, AZ 85719, USA} 

\affiliation{$^{5}$National Solar Observatory (NSO)/DKIST, retired, 3665 Discovery Drive, Boulder, CO 80303, USA} 

\affiliation{$^{6}$National Astronomical Observatory of Japan (NAOJ), 2-chōme-21-1 Ōsawa, Mitaka, Tokyo 181-8588, Japan}

\affiliation{$^{7}$Space Telescope Science Institute (STScI), 3700 San Martin Dr., Baltimore, MD 21218, USA}

\affiliation{$^{8}$Infrared Science Archive (IRSA), IPAC, MS 100-22, Caltech, 1200 E. California Blvd, Pasadena, CA 91125, USA}

\affiliation{$^{9}$IPAC, Mail Code 314-6, Caltech, 1200 E. California Blvd. Pasadena, CA 91125, USA}

\affiliation {$^{10}$ADS / Center for Astrophysics, Harvard \& Smithsonian, 60 Garden Street, Cambridge, MA 02138, USA}

\affiliation{$^{11}$National Radio Astronomy Observatory (NRAO), 520 Edgemont Road Charlottesville, VA 22903-2475, USA}

\thanks{$^*$E-mail: \href{mailto:jnovacescu@stsci.edu}{jnovacescu@stsci.edu}}

\begin{abstract}
Observatories need to measure and evaluate the scientific output and overall impact of their facilities. Observatory bibliographies are one of many valuable methods to assess impact. An observatory bibliography consists of the papers published using that observatory’s data, typically gathered by searching the major journals for relevant keywords. Recently, the volume of literature and methods by which the publications pool is evaluated have increased. Efficient and standardized procedures are necessary to assign meaningful metadata, enable user-friendly retrieval, and provide the opportunity to derive reports, statistics, and visualizations to impart a deeper understanding of the research output.

In 2021, a group of observatory bibliographers from around the world convened online to continue the discussions presented in \citet{2015ASPC..492...99L}. We worked to extract general guidelines from our experiences, techniques, and lessons learned. This paper explores the development, application, and current status of telescope bibliographies and future trends. The paper briefly describes the methodologies employed in constructing the databases, along with the various bibliometric techniques used to analyze and interpret them. We explain reasons for non-standardization and why it is essential for each observatory to identify metadata and metrics that are meaningful for them; caution the (over-)use of comparisons among various facilities that are, ultimately, not comparable through bibliometrics; and highlight the benefits of telescope bibliographies, both for researchers within the astronomical community and for stakeholders beyond the specific observatories. There is tremendous diversity in the ways bibliographers track publications and maintain databases, due to parameters such as resources (personnel, time, budget, IT capabilities), type of observatory, historical practices, and reporting requirements to funders and outside agencies. However, there are also common sets of Best Practices. This paper describes some of the results from our collaborative discussions.
\end{abstract}


\section{Introduction} \label{sec:intro}
It has become increasingly important for astronomical observatories to measure and evaluate the scientific output, overall impact, and general success of their facilities. Many methods exist to do so. A commonly used one is to create databases with descriptions of the scientific papers that have been published using data from the observatory facilities. The resulting compilations are typically referred to as telescope or observatory bibliographies and the technique of interpreting and analyzing them as bibliometrics. In many observatories, librarians and archive specialists manage these information resources. These staff also play a defining role in the further development of the bibliographies to ensure their continued value for the specific observatory including its management, governing bodies, and project and instrument scientists, as well as the wider astronomy community.

Observatory bibliographies may include telescope data publications, data products publications, data services publications, staff publications, mission bibliographies, archival products bibliographies, etc. They are typically compiled by scanning the major journals for scientific papers that use or analyze data generated by the respective observatories. During recent years, there has been a constant increase in not only the volume of astronomical literature \citep{2022PASP..134a4501C} but also the number and methods by which a pool of publications is evaluated \citep{10.1007/978-3-030-02511-3}. Efficient and standardized procedures are necessary to assign rich and meaningful metadata, enable user-friendly retrieval, and provide the opportunity to derive reports, statistics, and visualizations that enable a deeper understanding of the research output from various perspectives.

Astronomy has a long-standing history of sharing and exchanging papers, data, and software. Despite (or possibly because of) the large geographical distances between observatories, collaboration has always been a vital element to staying informed about recent developments and building on the results of peer researchers. This is equally true for astronomy librarians, with their series of meetings, such as the Library and Information Services in Astronomy (LISA) conferences\footnote{\url{http://www.eso.org/sci/libraries/lisa.html}}, and other formal and informal ways of exchanging best practices in a fair, transparent, and collaborative way. It is therefore not surprising that astronomy librarians, data scientists, and archive specialists have been discussing observatory bibliographies for many years. Many papers describing individual observatories’ bibliographies have been published (see Appendix \ref{sec:furthereading}), along with articles about Best Practices \citep{2015ASPC..492...99L,2022ApJS..260....5C}.

In 2021, a group of observatory bibliographers from around the world convened online to continue the discussions and efforts presented in \citet{2015ASPC..492...99L}. Many of the institutions represented in that paper are the same, although individuals have changed, and representatives from additional institutions were added to broaden the astronomy disciplines managing observatory bibliographies. We formed the Observatory Bibliographers Collaboration (OBC), with the goals of sharing transparent and fair bibliography methods and standards and creating shared documentation and processes. The group has worked to extract generally applicable guidelines from the experiences, techniques, and lessons learned of the various observatories.

Over the course of our discussions, a few key actions emerged that shape the objectives of this paper. We aim to:
\begin{itemize}
\item cover all aspects of observatory bibliographies to identify commonalities and to present the creative ways in which observatory bibliographers have met their challenges despite limited resources;
\item provide some lessons learned with a focus on the most serious errors over time and how they were addressed; and
\item provide updated standards/recommendations, recognizing that as science changes, so must observatory bibliographies.
\end{itemize}

The best practices that we outline in this paper are intended to serve as guidelines to both new and experienced bibliographers. We hope that we can provide guidance to those needing to create new observatory bibliographies because, as we discovered, a quick-and-dirty approach can lead to incomplete or wrong conclusions and may ultimately be worse than no bibliography at all. For those individuals tasked with maintaining observatory bibliographies who are not credentialed librarians, the descriptions and recommendations contained in this paper are intended to fill gaps in their knowledge of library practices. In addition, those individuals who use and assess the metrics generated from observatory bibliographies can gain an understanding of how these bibliographies are constructed and their constraints.

In this paper, we explore the development, application, and current status of observatory bibliographies and the direction in which they are moving. We briefly describe the methodologies employed in constructing the bibliographies, along with the various bibliometric techniques used to analyze and interpret the papers collected within these bibliographies. We explain the reasons for non-standardization, i.e., why it is essential for each observatory to identify the metadata and metrics that are meaningful for them; caution on the (over-)use of comparisons among various facilities that are, ultimately, not comparable through bibliometrics; and highlight the benefits of observatory bibliographies, both for researchers within the astronomical community and for stakeholders beyond the specific observatories. The paper concludes with a look at essential skills for observatory bibliographers, as well as a glimpse towards future trends and technologies that will influence the creation, maintenance, and further development of observatory bibliographies.

\section{Identifying Candidate Publications} \label{sec:candidate}

To begin the bibliography process, most bibliographers rely on automated and/or semi-manual keyword matching against the full-text files of publications provided by the publishers. Bibliographers then scan the literature visually and/or aided by text-mining software to identify candidate publications for inclusion in their observatory bibliography.

\subsection{SAO Astrophysics Data System (ADS)} \label{subsec:ADS}
The SAO Astrophysics Data System (ADS), funded by NASA, is a digital index operated by the Smithsonian Astrophysical Observatory (SAO) under NASA Cooperative Agreement 80NSSC21M00561\citep{2004A&G....45c...7E,2015ASPC..495..401C}. ADS currently contains more than 15 million records that cover the NASA Science Mission Directorate (SMD) disciplines of astrophysics, planetary science, heliophysics, and earth science. ADS will soon become the NASA Science Explorer (SciX), and any new bibliographic projects should be started on that platform when released. NASA-funded research in biological and physical sciences is also being added to SciX. Currently, the records are grouped among four primary collections: astronomy and astrophysics, physics, earth science, and general science. The bibcodes that ADS currently uses to identify publications are 19 digit, human-readable, unique identifiers that describe the article with a year, journal, volume, and page, e.g., 2022ApJS..260....5C. Many (though not all) of the observatory bibliography infrastructures are built around these bibcodes. Abstracts and full text of major astronomy and physics publications are indexed and searchable within ADS, and arXiv e-prints are ingested daily as well. The ADS website offers help pages as well as a blog and news updates.

Observatory bibliographers use the ADS database and its functionalities extensively to identify and track observatory publications. They can query the ADS database using either the ADS web submission search forms or the system’s application programming interface (API) to identify publications to be assessed for inclusion in observatory bibliographies. Users can search in a variety of fields, including full text, author, affiliation, keyword, publication name or abbreviation, and date. It is also possible to string together multiple search terms to develop a query. Users can create libraries of ADS records for specific telescopes, data products, missions, and other subjects. These libraries can be kept private, shared with specific collaborators, or made public for viewing by any ADS user. With assistance from ADS curators, many institutions have created publicly shared ADS libraries that can be set as unique bibliographic groups in ADS such as \textit{bibgroup:chandra} and \textit{bibgroup:GALEX}. Updates to these bibliographic groups can be set automatically or done manually on a quarterly, annual, or other regular cadence. Publicly accessible, curated bibliographies, which are datasets in themselves, are important components of open science and align with the current \textit{Transform to Open Science (TOPS)} initiative at NASA\footnote{\url{https://science.nasa.gov/open-science/}}.

\subsection{Crossref} \label{subsec:Crossref}
Crossref\footnote{\url{https://www.Crossref.org}} is a non-profit organization and aggregator for publication information. More than 17,000 organizational members and publishers deposit metadata about their journals, ebooks, conference proceedings, and other publications through Crossref. With the appropriate licenses and a Crossref agreement in place, an observatory can mine the full-text files for a set of journals, similar to full-text field searching in ADS. Crossref offers its own RestAPI\footnote{\url{https://www.Crossref.org/documentation/retrieve-metadata/rest-api/}}. We encourage those interested in processing large volumes of full-text files for bibliographic work to explore Crossref in addition to ADS.\\

\subsection{Keyword Searching } \label{subsec:keywords}
To comprehensively search through thousands of articles per year, bibliographers must consider all common forms of keywords, telescope names, instruments, high-profile programs, etc., and search acronyms, former names, accepted colloquial forms, and community-accepted spelling and punctuation variations in addition to the official name form. 

Comprehensive search and discovery using keywords are complicated by simple typos or misspellings not caught during the editorial process and differences in hyphenation and punctuation. While ADS does an excellent job of mapping synonymous terms and name forms in the background, they cannot be held responsible for identifying every variation, including newly created ones. As an example, the Two Micron All Sky Survey (2MASS) can also be found written as Two-Micron All Sky Survey, Two Micron All-Sky Survey, 2MASS, 2-MASS, and 2 MASS. Another example is the National Solar Observatory’s (NSO) Daniel K. Inouye Solar Telescope observatory, DKIST. It was previously called the Advanced Technology Solar Telescope (ATST), and in its beginning, the acronym did not include the first T, so was just AST. It is also referred to as Inouye, Inouye Solar Telescope, and DKI Solar Telescope. There are published papers using every one of these variations. 

For the Chandra X-ray Observatory, the keyword set includes Chandra, CXO, and AXAF and their variants. The European Southern Observatory (ESO) uses regular expressions to retrieve program IDs and data DOIs in papers and also searches for over 200 keywords; many of them are acronyms as well as long versions of facility and telescope names. Often acronyms reveal unrelated homonyms. An overview of the experience gained through more than 20 years of the ESO telescope bibliography is provided in \citet{2018SPIE10704E..0RG}.

Common acronyms are difficult to manage, and they add to the time needed to assess papers \citep{2022PASP..134e5001S}. In the case of HST and JWST, respectively, the acronyms for the Cosmic Origins Spectrograph (COS) and the Near Infrared Spectrograph (NIRSpec) were removed entirely from bibliography search strings due to the number of false hits. NIRSPEC (typically in capitalized form) is also used to identify an instrument on one of the telescopes at the W. M. Keck Observatory. Some mitigation strategies to deal with common keywords and acronyms in the literature include keyword pairs, such acronym 1 AND (keyword 1 OR keyword 2), e.g., NIRSpec AND (“James Webb Space Telescope” OR JWST), written parenthetically in ADS. Some acronyms can also be limited to certain journals. For example, a search can be created to ignore instances of the acronym ACS in \textit{SPIE Proceedings}, where ACS can refer to multiple concepts across sciences rather than its intended meaning – Advanced Camera for Surveys – more commonly found in astronomy journals.

NSF NOIRLab keywords include instruments, telescopes, surveys, the archive, and data products and services such as the Astro Data Lab. Searching for the Blanco and Mayall telescopes is complex, since individuals use a variety of names to refer to these long-lived telescopes. For example, the Victor R. Blanco Telescope is referred to as the Blanco, the CTIO 4m, the 4m telescope at CTIO, and the Cerro Tololo 4-meter/4m/4 m/4-m. In addition, the name Blanco appears often in reference lists and in star names so many false positives are retrieved when doing a search solely on the keyword Blanco. Similarly, the Legacy programs SIMPLE, GOALS, Taurus, Cygnus-X, which use large volumes of Spitzer Space Observatory data, were removed from the search terms due to the common use of these words in the literature \citep{2010SPIE.7737E..1VS}. 

For Spitzer, the keyword set included the short name of the observatory (Spitzer), acronyms for the instruments (IRAC, MIPS, IRS), and acronyms for the Legacy, Exploration Science, and Frontier Legacy programs. Spitzer has since been handed over to the Infrared Science Archive (IRSA), which searches on a subset of these terms. Searching by instrument names and acronyms in addition to telescope and observatory names is necessary because some papers may only cite the instrument without mentioning the observatory itself. For IRSA, it is not possible to track all papers using all datasets that have varying names and acronyms. Because of limited resources, only papers for certain datasets are tracked, typically the most popular such as WISE, Planck, 2MASS, and IRAS. Accounting for all of the papers that emerge from diverse datasets can be challenging, particularly when the publication does not specify the name of the observatory or even the instruments used to obtain the data.

\subsection{Alternatives to Keyword Searching and Limitations} \label{subsec:keyword-alt}
One might think that an alternative and more accurate search could be accomplished by looking for citations of key publications. The 2MASS project requests that people cite \citet{2006AJ....131.1163S}, the journal article describing 2MASS, but many people cite instead the delivery document, \citet{2003tmc..book.....C}. For Spitzer, only 62\% of papers that use data from the observatory cited one of the key publications about the observatory or used the acknowledgment statement correctly \citep[see][fig. 2]{2022PASP..134e5001S}.

Searching by a standard acknowledgement phrase to identify candidate papers is not a comprehensive solution, even when authors remember to include one. The 2MASS project has been around for nearly two decades and has been fundamental to many papers; the words in the requested acknowledgement seem to be “carried along” in the template LaTeX for many authors. There are many examples of papers that acknowledge 2MASS but do not appear to actually use 2MASS data, even for calibration, in the analysis presented therein. If those papers were not manually excluded by checking each one, they would artificially inflate publication statistics. Likewise, while it may be helpful to search the Acknowledgements section for mentions of grants, the existence of a grants statement does not guarantee that new, original analysis was done using that data. Counting instances of grant statements should not be relied on as a primary method to identify science papers attributed to a particular observatory. Still, the presence of a data acknowledgement statement can be a strong indicator to the bibliographer that they may be assessing a science/data paper.
\subsection{Communicating Data Citation Policies} \label{subsec:data-cite-policies}
Observatories are advised to clearly communicate the data citation policy(ies) relative to their observatory(ies). Such policies should state when and how authors must acknowledge use of the observatory’s data and provide standard verbiage that can be cut and pasted into a manuscript. Some suggestions for disseminating this information include posting on the observatory's public website, providing verbiage within calls for proposals, linking or re-posting on the landing page for the data archive, and sharing such policies in person-to-person communications with investigators and archive users. Some examples of data citation policies and suggested acknowledgments can be found on the IRSA, NAOJ, ESO, NOIRLab, Chandra, and STScI websites, as well as most other observatories.

\section{Evaluation of Publications for Inclusion in Observatory Bibliographies } \label{sec:inclusion}

After a candidate paper for inclusion is identified, the bibliographer analyzes the paper for inclusion in the observatory bibliography. Papers are commonly examined “by eye”, following the guidelines set out in this paper. Most papers are still tagged and categorized in a semi-manual fashion by bibliographers, requiring human review to determine whether publications that mention a telescope, observatory, instrument, data product, or survey in the full text do in fact qualify as reportable “science”, “data”, or “observational” papers.

\subsection{Science Papers} \label{subsec:science}
A critical component to enabling fair comparisons and benchmarking of observatory publications is a common definition for “science”, “data”, or “observational” papers. When a paper is classified as one of these, it is typically included in the official list of publications attributed to an observatory and tabulated in internal and external reporting metrics. Classification as “non-science” is discussed later in this section.

The OBC agrees on the following criteria for tagging papers we roughly categorize as “science”, otherwise called “data” or “observational” papers at some observatories. Each observatory may interpret these guidelines in slightly different ways, but the spirit of what defines a science publication is consistent across our observatories.

\begin{enumerate}
\item \textbf{To qualify as a science paper, it must be apparent that data or data product(s) from the observatory were used and that the data or data product(s) formed the basis for reaching a new scientific conclusion. Whether your observatory includes only publications that appear to use “raw” data or also publications that cite compiled catalogs and high-level products must be clarified within your organization and should be communicated in both internal- and external-facing metrics.}

Data and/or data products referenced in a publication can be from a single telescope or many observatories. Borrowing from the NOAO Inclusion Criteria document: “The amount of … data analyzed does not have a bearing upon inclusion; if a small amount of … data in relation to the total amount of data sets in the paper is used in the conclusions, the paper is included.” In another example, if the 2MASS data were being used to calibrate the astrometry and/or the flux for data obtained by authors at another observatory, and that other secondary observatory's data were used to reach a new scientific conclusion, the paper may still be included in the 2MASS/IRSA bibliography, even if the 2MASS data were not the basis for the new conclusion.

\item \textbf{“Data” can be defined in a number of ways – it may be original data taken by the author(s) as part of their proposal(s) or archival data reduced or analyzed anew by the author(s).}

Like the 2MASS example, the data might not be observational data but rather position or flux calibration for some observatories or datasets. Science papers may also rely on enhanced data products such as catalogs (including source catalogs) or spectra that combine processed telescope data in a meaningful format for the greater astronomical community, so long as it is apparent these refined data products were a component of the author’s analysis. Spitzer tagged all papers that used enhanced data products, which is useful in measuring the popularity of the advanced data products.

Bibliographers generally do not count publications that cite or acknowledge an observatory without actually having used its data. For many observatories, it is not uncommon to find a paper in which the author(s) acknowledge an observatory or one of its data products, e.g., 2MASS, HST grant number, or ESO project number, without actually using their data in the paper. 

\item \textbf{To qualify as a science paper, it does not matter if the observations have been published before, so long as the way the results are presented constitute new analysis or use.}

A simplified summary or quote mentioning how others used or analyzed data in an earlier publication generally does not constitute new use. A statement such as “Smith \& Doe found the same using the VLT” would not qualify, whereas a statement such as “see Bain et al. (2009), Table 3, for a list of VLT observations used in our analysis” would constitute reanalysis. 

\item \textbf{Further refinement of science papers is possible.}

For pointed observatories in particular, what constitutes a science paper depends 
on the mission-specific definitions of what constitutes “citable" science data. The types of scientific data available can evolve with time as the archive grows and the data lend themselves to scientific results based on higher-level processing and analysis. One goal of the mission bibliography is to expand and update the definitions of a “science paper" to keep step with the evolution of how data are used and the scientific results produced by the astronomical community during the lifetime of the mission and possibly beyond.

The Chandra X-ray Center (CXC) tags their science papers as either “direct” or “indirect”. In CXC’s definition, a Chandra Science Paper (CSP) is one in which Chandra data make a significant contribution to the scientific conclusions of the paper. “CSP-direct” papers start with data products that can be retrieved from the Chandra archive. “CSP-indirect” papers present new analysis of Chandra data starting from the results in a referenced paper or use high-level, science-ready data products derived from lower-level, archival observations. Indirect papers are predominantly based on catalogs comprised of Chandra single observations and the Chandra Source Catalog, which combines and processes single Chandra archival observations to provide access to science-ready properties and data products for all X-ray sources detected by Chandra. Both CSP-direct and CSP-indirect are included in the annual science paper statistics and qualify as science papers from the point of view of other observatories, even though other bibliographies may not separate “direct” and “indirect” science in the same way.

The Space Telescope Science Institute (STScI) and Hubble Mission Office decided to track the use of non-reprocessed HST data, in particular from catalogs, calibration studies, etc., in the literature and created a new paper classification called \textit{data-influenced} in 2018. These papers are not included in the annual science paper metrics since they do not present a (re-)analysis of HST data. They are significant to the Mission Office, however, because they present another use case for how the telescope enables astrophysical research, even if not counted among the official statistics. 

As the nature of astrophysical research changes, the policies and guidelines for inclusion of papers need to be updated regularly and adapted. As in the Chandra and HST examples, rules for the use of catalogs may have to be established to properly distinguish between papers that merely cite measurements or results from the literature and those that actually relied on downloaded (archival) data or (re-)analyzed them to achieve new scientific results.

\item \textbf{Each observatory makes the decision on whether refereed and non-refereed publications are counted as official science papers.}

Some observatories include non-refereed publications in their official science paper count while others do not. NSO includes SPIE non-refereed publications in their official count; the rest of the publications in their bibliography are refereed. STScI generally does not include non-refereed publications in its official science paper count, with the exception of some major conference proceedings (IAU, ASP, AIP, SPIE), constituting less than 4\% of HST science papers through 2021. The CXC bibliography ingests all data and catalog papers it finds in the worldwide literature, refereed or not. Chandra also includes some circulars (e.g., the Astronomer's Telegram) that supply the full content of the article as an abstract to the ADS. Those publications are considered full articles for the internal bibliography, but are not tabulated in external metric reports. Some proceedings publish only the abstract as the complete contents. Those publications are not considered to be full articles and are not entered into the CXC internal bibliography \citep{2012PASP..124..391R}. NOIRLab excludes all non-refereed publications in its official science paper count, as do Spitzer and IRSA.

Dissertations are another variable in science paper bibliographies. Some observatory bibliographers include them in reportable metrics, while others do not. In some cases, dissertations may be their own reportable metric, separate from the official bibliography. The inclusion of dissertations can be particularly important as they represent a metric (although indirect) of how missions impact younger generations of astronomers and help educate our community. Dissertations should be maintained as a separate category in the bibliography since often they represent, or will represent, scientific results later published in a refereed journal article. Not all dissertations are published in ADS so the bibliographer will need to mine other databases, such as ProQuest Dissertations \& Theses, for these publications.
\end{enumerate}

\subsection{Non-science Papers} \label{subsec:non-science}
The following list describes some common examples of attributes of papers that may be tagged as non-science publications, referred to as a “mention” by various observatories. This list is not exhaustive, but represents the most common scenarios when bibliographers may choose not to classify a paper as pure “science” or “data”.
\begin{itemize}
\item \textit{Lack of New Analysis}: Compares previously published results and scientific conclusions to results in current paper without new analysis. Uses only values extracted from observational data without new processing or analysis of the data. This includes instances where a paper uses measurements or results from the existing literature.
\item \textit{Background Content}: Cites observatories or missions in the history, background, or introduction of a research paper to summarize other people’s research or the entire body of literature.
\item \textit{Fields and Locations}: Uses a defined field first observed by the telescope to select where to point another telescope, without further discussion or proof that data from original observations in that field were analyzed beyond processed results in an earlier publication. Uses finalized results from an earlier publication to determine source selection and location of observations presented in the paper. Uses data to calibrate positions or fluxes for data from other telescopes, e.g., 2MASS.
\item \textit{Instrumentation and/or Software}: Describes instrumentation or software. Some observatories, such as NSO, include these papers in their official science paper count. Spitzer tracked these papers, but unless they used observational data from the observatory, they were not included in the official science paper counts. NOIRLab tracks instrumentation papers, but does not include them in their official science paper count, nor does STScI.
\item \textit{Archival Tools}: Describes tools or protocols at archives for astronomers to use. Some institutions, such as IRSA, may choose to include these papers in their official science paper count.
\item \textit{Future Observations}: Proposes future observations with telescopes or instruments by creating models or simulations, using data merely as examples or to validate a proposed model's output. These may be tabulated in a different way that is more significant than a passing “mention”, but are generally not counted as official “science” papers.
\item \textit{Images as References}: Shows images as a visual reference or as an overlay with other telescopes’ visuals without further discussion. This is typically done to layer observations from other parts of the spectrum or to indicate regions of the sky considered in a paper. A publication that uses visuals to achieve scientific results or provides additional details about visuals and their underlying data may qualify as a science publication.
\item \textit{Nested Catalogs}: For most observatories, only papers citing the original catalog and demonstrating use of the original catalog are included in the science papers. Papers citing a catalog that absorbed processed data from the original catalog are typically not included. Nested catalogs are common in extragalactic deep fields. To illustrate the subtleties, IRSA will count papers if they use data from sources that were downloaded from IRSA (or could have been downloaded from IRSA), including catalogs, but not products derived from catalogs, e.g., extinction maps, unless also served by IRSA. The Spitzer bibliography counted papers that explicitly stated they used the catalogs produced by the large survey Legacy or Exploration Science and Frontier Legacy programs, as those programs were supported by the observatory’s data analysis funding for the specific purpose of creating the catalogs.
\item \textit{Review Articles}: These generally summarize the literature rather than present new science results. Some may qualify as new research, but this is rare.
\item \textit{Advance Access Articles}: Advanced electronic versions of papers that later appear as a final publication are not included. This includes, for example, arXiv, MNRAS.tmp, and Natur.tmp bibcodes from ADS. These preprints may be tracked internally, but are not reported in external metrics as official science papers unless they later appear in refereed form.
\item \textit{Standard Stars}: Published standard stars for calibration purposes are generally not considered new science.
\item \textit{Enhanced Data Products}: Catalogs and other enhanced data products created by the community using data from programs not specifically funded to produce these high-level products. For example, on Spitzer, some programs resulted in the creation of enhanced data products, but are not official Legacy\footnote{\url{https://irsa.ipac.caltech.edu/data/SPITZER/docs/spitzermission/observingprograms/legacy/}} or Exploration Science\footnote{\url{https://irsa.ipac.caltech.edu/data/SPITZER/docs/spitzermission/observingprograms/es/}} programs. The Spitzer bibliographer did not count papers using these products because Spitzer did not necessarily host the enhanced products, and producing the enhanced products was not part of the funding that was explicitly allocated. However, IRSA does count these for some programs. The HST and JWST bibliographies tend to include papers that cite catalogs and other High-Level Science Products\footnote{\url{https://archive.stsci.edu/hlsp}} hosted by the Mikulski Archive for Space Telescopes (MAST) because part of MAST’s mission is to collect and curate community-contributed products based on MAST data to make them available to the wider community for reuse.
\end{itemize}

\section{Journal Coverage} \label{sec:journals}
Journal coverage is an important consideration in any discussion of bibliometric best practices. The set of journals searched impacts end counts for the number of science papers included in an observatory bibliography. Different institutions search different sets of journals to find science papers. Important factors in choosing the list of journals to cover include the historical publication record, the science areas covered by instruments or missions, and the resources available at an institution.

We tallied the representation of each ADS-indexed journal in our institutional bibliographies to gauge the historical record, reporting on the number of papers found in specific journals as a percentage of our total bibliographic dataset. The overwhelming majority of papers in every institution’s historic bibliography came from one of five core journals, and fewer than a dozen other titles are significantly represented. Among all the possible titles available in ADS, only a small subset is statistically significant to the various bibliographies maintained by the OBC members. The list of the most common journals across institutions is provided in Appendix \ref{sec:commonjournals}. The historical record makes a compelling argument for where to focus journal coverage in new or future bibliographies when staff and labor limitations dictate a curated approach from the start.

An important consideration in choosing journals to search is subject coverage. Certain fields in astronomy are inherently different from one another; for example, extragalactic versus solar astronomy or planetary science versus cosmology and may rely on different subdisciplines \citep{2019AAS...23345301H}. The journals that are relevant to these fields will vary. For example, \textit{Icarus} is better suited to planetary missions and their resulting publications, as are some American Geophysical Union (AGU) titles. Other journals such as \textit{Astronomy \& Astrophysics} are steady sources of new research on extragalactic and stellar topics. If an institution's instruments or missions are particularly focused on one area of practice within astronomy, this should inform the journals that a bibliographer searches for science papers. It is incumbent on a bibliographer to determine which journals are relevant for their observatory's data.

Bibliographers need to consider the resources available for their efforts – software, tools, staff time – when deciding which journals to search. The more journals included in a search and the more terms, keywords, or acronyms searched, the more results there will be for candidate papers. Tools, technology, and a considered approach to query strings will reduce the time required to review results.

There are three main strategies a bibliographer can employ when choosing which journals to search. The first is the broadest use of ADS where all available journals are searched, either using the API or manually running searches on the ADS site. The second strategy is to conduct a focused search on only a select list of journals, either through ADS or through the individual journals. The third strategy is a full-text search of specific journals using text mining or other technologies, which may rely on Crossref. A combination of strategies may be best depending on the specific circumstances. At times, there are benefits to casting a broad net to determine if data and discoveries by an observatory are impacting other disciplines in astronomy. Many observatory bibliographers use a curated approach throughout the year, then run a wider search against negative results at year’s end to identify relevant papers that may not have been found using the more precise criteria. One example is combining designated JWST keywords with \textit{NOT bibgroup:jwst} in ADS, scouring through additional refereed journals after known JWST papers have been tagged from the pre-defined journal set for the year.

The choice of journal coverage at an institution is specific to that institution's current needs, historical record, subject areas within astronomy, and available resources. Just as no two observatories or telescopes are identical, journal coverage that is the basis of an institution's bibliography will be different. It is critical that bibliographers and those who use curated bibliographies understand not all institutions are considering the same pools of journal coverage. If an institution beginning a new bibliography is unsure which journals to begin focusing on, we recommend considering the core list of journals highlighted in Appendix \ref{sec:commonjournals}. Other best practices in journal coverage we recommend are to use only refereed journals to build your count of science papers and to strive to attain a 90\% capture rate of those papers before considering non-refereed publications in ADS.

\section{Metadata} \label{sec:metadata}
Once a publication is chosen for inclusion in an institution’s bibliography, the bibliographer must assign metadata relevant to their observatory to that publication. In this section, we present guidelines and considerations for developing and applying metadata tags to publications in an observatory bibliography.

\subsection{Core Metadata Set} \label{subsec:metadata}
Below is a list of the core set of fields the OBC recommends that a new observatory considers collecting for local analysis. An existing observatory may consider using the recommendations to backfill their metadata for analysis and reporting, in order to develop a deeper understanding of science trends over time. This set is based on our shared experience of metadata that we 1) have found useful from the start and continue to find useful, 2) wish we had collected in the past and continued collecting, or 3) consider worth backfilling in our observatory bibliographies. Each observatory should make its own determination as to which of the core metadata fields are relevant for its purposes.

Bibliographers should regularly review this set of core metadata fields to keep up with publication practices and to identify changes in how the astronomical community uses data produced by their observatory. Expanding or refocusing the scope of bibliographic metadata collection is an unavoidable step when a new category of papers (covering, for example, new types of data products or software produced by an observatory) is being collected and classified.

Many of the recommended metadata fields are available in ADS, so documentation of them in an internal observatory bibliography may not be necessary. It is possible to analyze and retrieve these data from ADS without storing the metadata locally. Fields available in ADS are marked with an asterisk\text{*}.

\textbf{Identifiers}
\begin{itemize}
\item ADS bibcode\text{*}
\item proposal/program ID(s) or observation ID(s)
\item associated principal investigators (PIs) and co-investigators (Co-Is)
\item ORCID(s) for author(s), when available\text{*}
\item data product, dataset(s), or dataset(s) DOI(s); exactly how the data is bundled or cited will vary according to the data archive
\item software programs used in paper analysis
\end{itemize}

\textbf{Bibliographic Information}
\begin{itemize}
\item paper title\text{*}
\item publication/journal title\text{*}
\item publication date\text{*}
\item publication volume, issue, and pages\text{*}
\item published article DOI (version of record)\text{*}
\item optional: preprint DOI or identifier\text{*}
\end{itemize}

\textbf{Author Information}
\begin{itemize}
\item first author name*
\item first author affiliation(s)* 
\item first author country/ies*
\item additional author name(s)*
\item additional author(s) affiliation(s)* 
\item additional author(s) country/ies*
\item observatory staff publication (if tracked by your institution)
\item author(s) category/ies (e.g., postdoc, visitor, student) 
\end{itemize}

\textbf{Publication Type}
\begin{itemize}
\item refereed / non-refereed* (if including both types in bibliography)
\item thesis/dissertation (often, but not always in ADS)*
\item paper classification: science, non-science/mention, data-influenced, instrument, engineering, etc.
\end{itemize}

\textbf{Telescope and Instrumentation}
\begin{itemize}
\item telescope/observatory (for bibliographies that capture more than one telescope’s data)
\item instrument(s), mode(s), channel(s), etc.
\item filters(s) and wavelength(s)
\end{itemize}

\textbf{Data Products and Services}
\begin{itemize}
\item archival flag (PI/observer vs. archival use)
\item observatory data services or high-level products used (for example, NOIRLab’s Astro Data Lab, or the Chandra Source Catalog)
\end{itemize}

\textbf{Surveys/Missions}
\begin{itemize}
\item survey observatory or pointed / object-specific observatory
\end{itemize}

\subsection{Additional Metadata for Inclusion} \label{subsec:add-metadata}
Each observatory has unique characteristics, needs, and reporting requirements, which may necessitate the inclusion of additional metadata fields. As examples, the following metadata fields may be considered for inclusion: 
\begin{itemize}
\item Page charges paid by observatory
\item Data analysis funding/grant information
\item Length of time to publish since data were obtained or released
\item Names of other observatories whose data were used (for multi-mission papers)
\end{itemize}

\section{ADS Bibgroups} \label{subsec:ADS-bibgroups}
ADS offers bibliographers the option to create bibliographic groups (called \textit{bibgroups} in the ADS web submission form and API) to collate publications that used their observatory’s data. Bibgroups also make it easier to identify multimission/multiwavelength papers that use data from multiple telescopes or observatories. The Union, Intersection, and Difference functions for ADS Libraries\footnote{\url{https://ui.adsabs.harvard.edu/help/libraries/}} can augment discovery and arrangement of new paper sets focused on two or more observatories. Some real-life examples include searching for papers on a particular scientific topic, e.g., proper motions and astrometry for stellar clusters using HST data only, Gaia data only, or combined HST and Gaia data. A researcher planning a new proposal focused on AGNs, quasars, or neutron stars might also use the \textit{bibgroup:chandra} in combination with \textit{bibgroup:NOIRLab} to understand what past research has been done on their intended target(s).

For new bibgroups, ADS recommends that you create an ADS account for your group to manage the bibliography. While logged into this account, execute your search. Then proceed to add your results to a private library and set it to public. These libraries can be shared with other team members, and they will contain the canonical bibcodes for the publications contained in them. Bibliographers can maintain multiple libraries and use powerful tools to generate one master library from the individual libraries or provide the ADS team with all individual libraries to be combined into one bibgroup. After you have created the library or libraries for your bibliography, send the URL(s) to adshelp@cfa.harvard.edu. ADS staff will formally create the bibgroup in ADS. From this point on, you can manage your bibliography directly by adding or removing items to the related library(ies), and ADS users will have access to the most recent updates.

For the ADS libraries and bibgroups to be useful, each observatory must maintain the contents of its own bibgroup(s) and provide ADS with the parameters of its bibliography when first setting up a new bibgroup. Moreover, maintaining one bibgroup per telescope rather than per organization makes combined searches easier to execute. Ideally, bibgroups for active observatories are maintained through the most recently completed calendar year, or the one prior, so that more complex searches are possible.

\section{Reports and Metrics} \label{subsec:reports}
Publication lists and metrics are valuable data that are often included in reports to funding agencies, governing bodies, and the astronomical community to demonstrate the scientific output and value of the observatory. An observatory’s annual report typically includes this information; in some cases, partial information is included in monthly and/or quarterly reports as well. 

In addition, observatory bibliographers are often asked to provide publication counts and lists on-demand for a specific topic or a subset of programs, as required by research staff, funders, and other stakeholders. A recent example from STScI is an internal assessment of the number and types of papers using ULLYSES legacy program\footnote{\url{https://archive.stsci.edu/hlsp/ullyses}} data, gathered by trained bibliographers as opposed to publications loosely associated with the single keyword ULLYSES in an ADS full-text search.

Beyond ADS bibgroups, graphical representations of publication metrics are useful additions to observatory websites. For example, NRAO maintains online graphs that show refereed paper totals by instrument and year, as well as citation totals by year for the most recent 10 years. Chandra maintains online graphs showing the total number of observational papers per year and links the tables used to create the plots, both in total and broken down by metrics such as quarter, category of target, grant type, etc. NOIRLab has a dynamic publications dashboard that shows publication counts for each of their telescopes for the past 10 years which links to its ADS public libraries for individual telescopes and data products. The public interface of ESO’s telescope bibliography provides a large variety of options to export and visualize search results. ESO librarians combine the most frequently requested reports and publish them on the web\footnote{\url{https://www.eso.org/libraries/pubstats}}. STScI provides visual representations as well as public-facing numeric counts through the previous calendar year for both HST and JWST.

There are risks attached to reporting on metrics, especially when trying to interpret those maintained by another observatory. See section \ref{sec:caveats} for a discussion on why comparing publication numbers among observatories is problematic. It is vital that each bibliographer makes the Criteria for Inclusion evident so that the metrics can be evaluated appropriately.

\section{Caveats of Comparing Bibliographies} \label{sec:caveats}

\subsection{Intra-observatory Comparison} \label{subsec:intra-obs}
An intra-observatory comparison is any comparative assessment for the period of time during which the observatory has been operating, for example, looking at the number and types of science publications using Spitzer data before and after the start of its warm mission\footnote{\url{https://www.jpl.nasa.gov/news/nasas-spitzer-begins-warm-mission}}.

Often, technical events can visibly modify the bibliographic landscape for a period of time. As an example, one of the Chandra instruments – the High Resolution Camera (HRC) – stopped functioning in February 2022. The engineering team managed to bring it back to life in April 2023\footnote{\url{https://groups.google.com/a/cfa.harvard.edu/g/chandra-announce/c/Z_ggkSlEV2w}}. Without this background information, someone could interpret the reduction in number of papers using HRC data during and following the HRC hiatus as a decrease in the scientific utility or interest in the detector, due to the amount of time it takes from data collection to publication. In reality, the dip is caused by the fact that no HRC observations were being taken for approximately 14 months. The CXC bibliographers are anticipating a statistically significant increase in the number of papers using HRC sometime in the future (2024--2025), as a result of the larger number of HRC observations (relative to normal scheduling) that have been taken by Chandra since resumption of HRC operations. Without context and an understanding of observatory operations, the dip and peak could be misinterpreted as being real ``bibliographic" phenomena when instead these extremes often reflect technical issues experienced by a facility.

Similarly, the HST saw declines in the total number of data publications in the periods of time leading up to and sometimes following the various servicing missions\footnote{\url{https://science.nasa.gov/mission/hubble/observatory/missions-to-hubble/}}. There was a notable decline in 1998--1999 (Servicing Mission 3A) and again in 2008--2009 (Servicing Mission 4), shown clearly in the annual HST Publication Statistics\footnote{\url{ https://archive.stsci.edu/hst/bibliography/pubstat.html}}. The number of papers that rely on data obtained using a particular instrument(s) is closely tied to each servicing mission and the suite of instruments installed or decommissioned during each mission. This in turn impacts the type of archival observations to which the community has access as instruments are placed online or taken offline. In short, observatory bibliographies are never independent of what is happening on the ground or in space and are closely linked to the operational status of the facility. 

Tracking the use of purely or partially archival data is another common use of intra-observatory comparisons. By tracking the gradual increase in archival data use, it is possible to advocate for ongoing support and funding for the data archive while the community continues to access and rely on that data to make new scientific discoveries, often in combination with fresh guest observer data or other observatories’ data. Having an archive seems to double the amount of articles produced by an observatory \citep{2022PASP..134e5001S}, something that is known specifically because of publication tracking.

\subsection{Inter-observatory Comparison} \label{subsec:inter-obs}
It is difficult to compare bibliographies between observatories, even ones that seem like they might be comparable. As discussed in sections \ref{subsec:keywords} (Keywords) and \ref{sec:journals} (Journals), the search criteria applied by each observatory can also have an outsized impact on the bibliography, making it difficult to conduct true comparisons between different facilities. Below we identify other issues the OBC members have encountered when asked to complete inter-observatory comparisons and our collected experiences trying to fairly compare observatories when required to do so.

Some journals do not update all of their article metadata or full-text files for up to three months, which makes searching for candidate papers difficult until the following quarter. Therefore, one cannot compare the yield of articles in 2024 until April 2025 at the earliest, for example, if a comprehensive annual bibliography is expected. It is important to ask: “Is the information I have, especially from outside my observatory, up-to-date for the investigation I have in mind?”. While it is possible to search and limit to a specific observatory’s bibliography using the ADS bibgroups, and then limit by refereed/non-refereed status as needed, generating equitable comparisons can prove problematic. All too often well-meaning staff use partial or incomplete data made available through an ADS bibgroup or public-facing page for a different institution and falsely assume it is “current” through the year or month for which the assessment is being done. This has resulted in uninformed, inaccurate, and misleading comparisons across institutions and observatories. Due to staffing and the manual nature of bibliographic work at present, many bibliographies run six months to multiple years behind in assessment.

For this reason, the OBC recommends contacting the bibliography staff at peer institutions whenever stakeholders or funders request an inter-observatory comparison. By understanding the limitations of your own publications database and those of peer institutions, you can more clearly state the caveats attached to the comparative metrics you provide and do your best to show a like-for-like comparison. 

In its overview of basic publication statistics, ESO provides a number of prepared statistics and graphs about their facilities’ science papers. One of them (“ESO and Other Observatories”) puts ESO’s research output in context by showing publication numbers from other observatories. However, ESO explicitly points to the caveats arising from such a comparison. Comparing the numbers of publications is the most simplistic way of comparing facilities, and it favors large institutions with many telescopes and instruments over smaller ones. A more meaningful investigation would normalize the numbers in some way, e.g., by number of observing hours, by actual share of data used in the papers (since many scientific articles use data from more than one observatory), or by budget. The “ESO and Other Observatories” section also states that when comparing publication statistics among different observatories, it is essential to assess the selection criteria applied by each observatory and that comparison data and graphs have to be used with utmost care.

An example of when cross-observatory metrics can be useful, even in broad stroke comparisons, is with regards to larger societal or economic trends. In 2020 and 2022, the Basic Publication Statistics maintained in ESO’s telbib database showed a marked decline in both 2020 and 2022 across multiple observatories, which correlates with the initial shutdown of facilities in early 2020 and the transition back to on-site operations through 2022 during the COVID-19 pandemic. While such metrics can rarely demonstrate causation, they do give stakeholders pause when trying to understand observatory operations in the wider context beyond the astrophysics community. 

\subsubsection{Survey vs. pointed missions} \label{subsubsec:inter-obs}
When proposing surveys that require many hours of observations, astronomers need to make a strong case that the survey data will be useful enough to warrant the allocation. This tends to result in programs that a broader range of astronomers support and use and data that are used for other purposes long past when they were taken.

For example, for Spitzer, the Legacy programs were large, coherent investigations proposed prior to launch and carried out early in the mission. The teams were funded prior to launch specifically to enable both rapid data reduction and delivery of high-level data products back to IRSA\footnote{\url{ https://irsa.ipac.caltech.edu}}. The surveys that were done in the first 2.5 years of the mission produced a much higher “papers per hour of data” rate (0.37) than the non-survey General Observer observations (0.17) taken at the same time --
see fig.~9 in \citet{2022PASP..134e5001S}. As time passes, the survey programs from that observatory continue to provide most of the papers as observations from those survey programs are republished \citep[fig.~6a and 10]{2022PASP..134e5001S}. The Legacy programs were recognized to be so successful that the community continued to propose large investigations that promised deliveries back to IRSA. These ready-to-use data products continue to generate substantial use of Spitzer data and citations. 

All-sky surveys such as Gaia, 2MASS, or (NEO)WISE can be used for nearly any investigation, provided the targets are bright enough. Observatories that are primarily focused on pointed observations (such as Chandra, Hubble, Spitzer, or JWST) will not necessarily have data at any given location on the sky. By their very nature, then, all-sky surveys have the potential to generate more publications and citations than pointed observatories.

\subsubsection{Large collaboration papers vs. small research groups} \label{subsubsec:icollabVsmall}

Missions that started with large international collaborations will often result in more papers referencing original collaborative publications and may accrue more citations compared to observatories that traditionally have smaller teams observing and using data.

Similarly, Guaranteed Time Observations (GTO) programs often result in proportionally more papers, since the teams of researchers are funded over a longer time span, often lasting for two or more years/cycles. The influence of accumulated volumes of data on observatory bibliographies is discussed further in \citet{2005ASPC..347..380B} for Chandra and shown in fig. 9 of \citet{2022PASP..134e5001S} with regards to Spitzer.

\subsubsection{Science topics and community size} \label{subsubsec:topicsize}
Observations focused on topics that are less widely studied or have smaller research communities, such as masers, or specialized methods or topics, like IR spectroscopy of Galilean satellites, will inherently have smaller bibliographic counts. Observations of targets that are more popular in the public sphere or have been observed for a longer period of time, e.g., M31, can result in a greater number of publications and more citations. Searching for the object M31 in the HST bibgroup using ADS yields over 1,800 results, accounting for more than 12\% of the entire HST bibliography to date.

Even just within the Spitzer archive, which should be similar in terms of people interested in the wavelength regime and consequent astrophysics, the spectra from IRS have a lower publication per hour rate than the two imaging instruments, IRAC and MIPS \citep{2022PASP..134e5001S}. Some of that likely arises from the fact that an image can serendipitously capture objects whereas all the spectroscopy must be deliberately pointed, but it is also likely that more specialized knowledge is required to analyze the spectroscopy, and therefore intrinsically fewer people attempt it. 

Spitzer also provides a ready-made lesson on the change in publication rate per observing hour when it comes to the nature of the observations. Early in the mission, many of the large time allocations were given to projects covering relatively large areas of sky. Those projects’ data continue to accumulate myriad citations as scientists use data from those surveys in conjunction with more recent observations. Later in the mission, after the cryogen was exhausted and only the shortest two channels of IRAC were still working, a much higher fraction of the observing time was devoted to projects that were monitoring targets for time variability – that is, staring at one place on the sky. Once a paper is written presenting the light curve (brightness as a function of time), future investigators cite that paper, rather than reusing the original data. This results in far lower publications per hour of observing time, given the strict definition for science papers; see further discussion in \citet{2022PASP..134e5001S}.

\section{Technology Considerations \& Standard Tools} \label{sec:tech}

An automated tool for searching journals or text mining can substantially reduce the number of staff hours required to review a list of journals for candidate papers. Questions to ask about opportunities to apply technologies at your institution should include the following:
\begin{itemize}
\item What expertise is available to code and curate tools to ensure they remain effective for paper discovery? 
\item Is it better to a) search journal tables of contents, b) search within full-text papers using the ADS GUI, or c) use an API such as those from ADS or Crossref to search journals’ full-text content? In the event a publication falls outside of the ADS' current holdings and there is a strong incentive for inclusion on behalf of the astrophysics community, the bibliographer should reach out to the ADS to request it be included.
\item Are all the necessary publisher licenses in place to allow access to papers in full for text-mining and detailed evaluation, whether
using ADS, Crossref, publisher web forms, or APIs?
\item How should my entry point(s) for full-text searching inform the design of my automation and tools, including the program or database in which I track verified science papers?
\item How will I retain a list of papers I consciously ignored or keep track of which journals and volumes I have already evaluated? 
\end{itemize}

Beyond the search tools and search methodology, you will need to ask yourself the following:
\begin{itemize}
\item What additional metadata, beyond what is available in ADS, does my observatory wish to maintain alongside each publication record?
\item Do I need to include publications that fall outside ADS, and if so, how will I ingest metadata from those publications, and is the metadata comparable to what is found in the bulk of the literature in ADS?
\item How will paper categorization affect design? For example, if tracking non-science papers in addition to pure science papers, should those non-science papers be split between historical references and forward-facing simulations?
\item How often will my stakeholders want reports on publication metrics and how might that influence design and processes? Which metadata fields and values will be necessary to track to satisfy reporting needs? See section \ref{sec:metadata}, Metadata, for further discussion.
\item Am I (or the appropriate software development team) expected to build a public-facing component specific to my observatory in addition to an ADS library or bibgroup? If so, how will it be updated from the internal bibliography tool(s)?
\item Am I (or the appropriate software development team) prepared to maintain the database and continually develop it in response to outside systems such as ADS and Crossref?
\item Am I (or the appropriate software development team) prepared to maintain ongoing documentation and versioning of the observatory bibliography software?
\end{itemize}

\subsection{Standard Tools: Low to High Investment} \label{subsec:tools}
Although many commercial products exist for managing bibliographies, no known off-the-shelf software exists for building or maintaining an observatory bibliography, and adopting another institution's application frequently requires a great deal of modification and technical expertise to fit your observatory’s needs. To manage a sustainable bibliography, the bibliographer (or your institution) may need to create in-house tools to get the job done. Here we present multiple options, from basic, low-investment spreadsheets and ADS libraries to complex databases integrated with an institution’s data archive. 

Some observatories may find they need to rely on a basic approach because of restrictions on resources and IT support. Varying levels of IT support, limited access to software developers within your organization, time, and the bibliographer’s skills and training may all impact design decisions when creating a new bibliography. It is important to bear in mind that without dedicated and ongoing support, observatories cannot easily add complexity, adjust their metadata model and paper categorization, or make significant changes to the software underpinning the database.

A perfectly usable, low-investment option is a basic MS Excel workbook or spreadsheets, tracking the ADS bibcodes or other unique identifier of papers that qualify as science and non-science papers (if also tracked) that are associated with your observatory. A related ADS library and bibgroup can be established and updated in tandem or at regular intervals. This is the approach used by NOIRLab and is how most observatory bibliographies begin. In addition, NOIRLab built a web interface for bibliography metrics, which pulls directly from ADS.  

The DKIST Observatory bibliography at NSO is maintained in an MS Access database. Other observatories, such NRAO, use a slightly more complex, homegrown multi-table database. NRAO uses a browser-based middleware application to interface with the data stored in a MySQL database, which reduces the need for library staff to have expertise in SQL and performs simultaneous transactions more efficiently than running multi-table queries. MySQL is used most often for queries to gather metrics requested by staff. NRAO also operates a faceted search tool running on Blacklight and Solr that the public can use to search for papers in the bibliography. Two dynamic graphs produced by a separate API show paper totals and historic citation totals.

Spitzer made an upfront investment in its bibliography early in the mission’s lifetime. Until it was decommissioned in 2020 and the data transferred to IRSA, Spitzer used a MySQL database with Shell and Perl scripts to interact with the database and three tables for pending, valid, and rejected papers. For more information see \citet{{2010SPIE.7737E..1VS}}. The Spitzer database did not include an internal general user interface (GUI) like some other observatory bibliography databases, but did allow the bibliographer to associate program IDs with papers. Tables were used to populate the \textit{Spitzer Biographical Database} web search interface which was public facing.

Other observatories have invested more time or software development resources in recent years and manage observatory bibliography software that is highly complex, with multiple relationships and dependencies between ADS and their institution’s data archive. Examples of more complex bibliography tools can be found at the Chandra X-ray Center, ESO, and STScI, all of which offer a UI for bibliographers who are not adept at software engineering in their day-to-day work. All three link the associated data from their respective archives to publication records in ADS.

ESO’s telescope bibliography (telbib) started in 1995. Its predecessor was a simple list of observatory papers compiled for the Annual Report. Since then, telbib has evolved into a sophisticated system with many features and facets. For the current system, ESO has developed a web application called FUSE (a fulltext search tool using PHP/MySQL) that queries ADS via their API in order to identify relevant papers. The telbib software is managed through another web interface (PHP/Sybase). It allows rich tagging with metadata, a large variety of export options, and reports in various formats. In addition to these internal applications, there is a public interface (PHP/Solr) that allows searching, exporting, and visualizing data via a web interface or an API. 

The Chandra bibliography is managed through a graphical user interface written in Perl and SQL and is accessible through a web interface. Due to the level of customization, especially for managing metadata and connections between CXC systems, it may be difficult to revise the bibliography software for other missions in future. In hindsight, the CXC bibliographer team stated it would have designed a database with less technical complexity so it is more adaptable for other Center for Astrophysics (CfA) needs. 

The STScI and MAST bibliographies are managed through a front-end graphical user interface used daily by archive scientists and bibliographers, with underlying code rendered in Python and SQL. The local tool is called PaperTrack, and it was rebuilt in 2018-2019 to take full advantage of the ADS API released alongside the ADS 2.0 platform.

PaperTrack is used for three main purposes. PaperTrack stores bibliographic information about publications found in ADS that use mission data (Science), are related to mission engineering or instruments (Engineering/Instrument), or reference MAST missions/data already published (Mention/Supermention/Data-Influenced). Secondarily, the tool provides institutional benchmarking numbers to STScI; its mission offices; NASA; its parent organization, AURA; and external bodies. Annual counts of refereed science publications are reported as one measure of scientific output. Lastly, STScI commits to an annual update for its mission metrics in late spring of each year in order to provide other observatories and science centers a record of publications that use or reference MAST mission data for their own benchmarking measures. This last use is discussed in more detail in sections \ref{subsec:reports}, Reports \& Metrics, and \ref{sec:caveats}, Caveats of Comparing Bibliographies.

To identify papers, STScI relies on the ADS API to look for relevant keywords in the full-text files housed in ADS. Not every publisher deposits its full-text files at a regular cadence with ADS, which limits the utility of the ADS API, through no fault of the ADS team. Candidate papers are manually assessed to determine if they only mention one of its three flagship missions in passing, or clearly used data to reach a new scientific conclusion. The HST, JWST, and the Nancy Grace Roman Space Telescope (RST; NGRST) comprise the flagship mission searches. The same assessment occurs for over 15 MAST missions including TESS, GALEX, Kepler, IUE, and many others. Searching the literature for candidate papers using MAST missions’ data is managed by the MAST archive scientists. MAST offers a public facing, external database, but for reasons expressed in section \ref{sec:caveats}, Caveats of Comparing Bibliographies, it is always advisable to contact the STScI bibliographers for assistance gathering the metrics you need. The database can be accessed at \url{https://archive.stsci.edu/publishing/bibliography#/}. It allows external users to view information about mission publications for previous years, find papers using their HST or JWST program data, and gather basic publication metrics information.

When designing bibliography software or choosing to implement a low-tech option, consider to what degree you are willing to rely on other technology. Scraping full-text html, for example, even if permitted by the publisher, will require intervention each time a publisher’s journal platform changes. If the ADS API undergoes a major re-haul, it may break existing scripts that allow you to search full-text files and ingest candidate paper metadata into your database. Consider reducing unnecessary dependencies on other systems whenever possible to ensure long-term maintenance. It is critical that those involved in building new bibliography software or enhancing existing software communicate proactively with the groups who manage related systems, such as your institution’s data archive, ADS, and publishers.  

\section{Staffing} \label{sec:staff}
An observatory bibliographer works with a diverse group of technical, scientific, and non-technical personnel both within the observatory and at outside research institutions, funding agencies, publishers, and observatories. Strong communication skills, a knowledge of the scientific literature, and proficiency in database searching are all important competencies for observatory bibliographers. Many bibliographers are either the sole staff or part of a small team performing these tasks, so a commitment to professional networking is a valuable asset.

One of the advantages of having dedicated staff is continuity and a homogeneity for classification and applied metadata across large sets of papers. While researchers and archive scientists may comprise the entire classification team, or be a key part of it, they may not be immediately comfortable with or skilled in classification. The task of classifying papers for mission bibliographies requires significant training and specialization that cannot be improvised. Archive scientists and research staff may understand which datasets and observations were used in papers, but the way a scientist reads and understands a paper is different from what is required of a classifier, though the same individual may serve in both roles with adequate training. Documentation outlining the various paper categories, metadata fields captured in the bibliography, and the process for evaluating a paper from a bibliographic standpoint are imperative. 

An observatory bibliographer’s roles and responsibilities typically include the following:
\begin{itemize}
\item Searching the literature to identify papers that use data from an observatory’s telescope(s) or that were published by observatory staff
\item Reviewing the literature for inclusion in the observatory bibliography, according to the observatory’s criteria
\item Tagging relevant publications with the appropriate metadata in the database(s)
\item Investigating and solving problems related to metadata, linked archival data, and ADS index records
\item Updating and improving data citation and metadata standards on a continual basis
\item Educating users on how to use and interpret bibliography tools and reports
\item Explaining the publication tracking process to bibliography stakeholders and observatory staff
\item Interpreting publication metrics and preparing reports that present numeric and visual data for both recurring and ad hoc needs 
\end{itemize}

\section{Future Trends \& Technologies} \label{sec:future}
Efforts are underway to implement an automated paper classification system at STScI/MAST to identify science papers within a set of candidate papers; however, even if this product comes to fruition in the 2020s, it is expected that human intervention will be needed to extract additional information about each paper to complete certain metadata fields such as program ID(s), instrument, and observing mode (for JWST). Additional work to understand how this system will handle false hits on papers that contain keywords or phrases of interest, but have nothing to do with MAST or flagship missions, is underway. A summary of the pilot, its accuracy and completeness, and remaining challenges is expected to be released in 2024 (Pegues et al., in progress). The first alpha release for the automated classifier is anticipated in 2025. It is expected that only a portion of the papers can be confidently tagged as science papers. A sizable portion may still require human review to accurately assign a paper classification such as science, mention, engineering, etc. Moreover, not all publishers are willing to provide access to entire full-text files, even for nonprofit, internal, non-generative purposes; the automation tool might only be possible for use with content from certain publishers until licensing agreements can be reached to allow for use of additional content in machine learning (ML) tools. Licensing restrictions generally apply to full text obtained via the publisher’s online platform in html or pdf form, as well as the full-text JSON files available through Crossref.

Similar efforts are also taking place at other observatories. For instance, an ML approach to identify data in scientific papers for inclusion into NED (the NASA/IPAC Extragalactic Database) has recently been described by \citet{2022PASP..134a4501C}. Similar to STScI and MAST’s experience, staff at NED expect that their ML tool will provide a partial solution but are still seeing a return on investment: “This ML application on both subject and content classifications has been in operations since January 2021, and has gained the team about 0.4 FTE, i.e., the equivalent of 4 person-years of labor over a decade of NED operations”. At ESO, an AI interest group is using the observatory bibliography as an example to study how algorithms can enhance the process of properly classifying research articles. 

While all of these efforts hold promise for more efficient, complete, and timely paper classification and metadata extraction, it will be essential to keep a balanced view of the role of new and emerging technologies in observatory bibliographies vis-a-vis human control. Per \citet{2022PASP..134a4501C}: “It is worth noting that this ML approach does not completely remove human involvement in the process. Human expertise and learning are needed for marginal cases that are not resolved by existing capabilities. Further, the trained model is applied to a domain-shifted dataset for predictions. Newly-emerged data content, variations in formats and representations can all contribute to deteriorating performance over time. Hence, there needs to be a continuing education program for retraining and updating the classifier models with new literature, which will require new labels identified by human experts periodically”.

\section{Lessons Learned} \label{sec:lessons}
Over the course of our discussions, we have formulated some lessons learned and guidelines to help others understand the work that goes into observatory bibliographies and their purpose. We also anticipate this article will help those starting new bibliographies, as well as those interested in revising current ones. Bibliographies provide value to their institutions, individual researchers, and data archives, as well as oversight committees and funding agencies.
\begin{itemize}
\item \textbf{To Institutions}, they demonstrate scientific output, provide data on telescope/instrument/archive use, and act as templates for either new or revised observatory bibliographies. 
\item \textbf{To Researchers}, they show their research teams’ impact and reach, aid in understanding how their work fits within the astronomical community, help identify areas of interest and potential collaborations for ongoing work, lead to suggestions for new facilities and missions, and support current projects and upcoming proposals. Researchers use observatory bibliographies to enable mission and/or data archive-oriented searching in ADS by combining the fields for \textit{bibgroup}, \textit{data}, and/or \textit{object}. The search results may reveal research on an object or class of objects previously observed with one or more facilities or wavelengths. For example, a researcher can conduct the search \textit{object:hd22879 data:eso} to find and then assess earlier observations when writing a new proposal to ensure limited duplication and to better inform their proposal aims. Another real-life example of how researchers have used curated mission bibliographies is when updating a core exoplanet textbook. A search across the whole of the JWST bibgroup helped the authors identify and classify exoplanet research enabled by the observatory to date.
\item \textbf{To Archives}, they demonstrate connections between publications and data usage over time, especially for archival data use. Many bibliographies enrich the ADS records with links to datasets or DOIs, making it easier for the community to locate data used in papers which leads to reuse, validation of research findings, proper citation of existing data(sets), and new discoveries. Regardless of the tool your institution develops, the OBC advises new bibliographers to manage the flow of new candidate papers and the steps involved in paper assessment on a rolling basis. Doing so will ensure connections between your bibliography software, ADS, your institution’s data archive, and public systems are functioning as expected at regular intervals.
\item \textbf{To Oversight Committees and Funding Agencies}, they demonstrate utility and level of use of facilities, observatories, and archives by the astronomical community; show return on investments; inform support priorities; illustrate how observatories are advancing science; and guide decisions for future facilities and archives development, especially through reports to Visiting Committees, User Groups, and Decadal Survey efforts.
\end{itemize}

Some common insights and shared considerations across observatory bibliographies are as follows:
\begin{itemize}
\item \textbf{ADS: }ADS (soon to be SciX) is an important tool for searching the scientific literature, deriving metrics, and maintaining public bibliographies.
\item \textbf{Criteria for Inclusion:} Each institution should document their Criteria for Inclusion and make these publicly available to guide bibliographic work and inform users of the bibliography and its parameters. Institutions and bibliographers need to evaluate and revise their bibliographic policies and procedures regularly to account for changes in areas such as reporting requirements, missions, goals, keywords, and technology. There is tremendous diversity among bibliographies related to the unique characteristics, goals, and subject area focus of each observatory. The criteria should clearly explain the observatory’s preferences and rationale for what is captured in their bibliography.
\item \textbf{Science Publications:} Most bibliographies focus on refereed “science” papers. A “science” paper is one where 1) data or data product(s) from the observatory were used, and 2) the data or data product(s) formed the basis for reaching a new scientific conclusion. It does not matter if the observations have been published before, so long as the way the results are presented constitute new analysis or use. 
\item \textbf{Journal Coverage: }There is a core subset of journals that most observatories search to compile their bibliographies. Differences in journals searched beyond this subset relate to the unique characteristics, capabilities, and science goals of the observatory.
\item \textbf{Metadata:} Bibliographers need to regularly assess their metadata tags as these will change over time as science evolves and the institution transforms. Having a bibliography from the start will allow the observatory to identify which papers it may need to reclassify or augment through a more robust classification scheme in the future. Observatories should consider which data points to collect from science publications from first light/launch onwards to demonstrate to funding agencies that the observatory is making a difference in the research field(s) and enabling new findings.
\item \textbf{Comparing Bibliographies:} Direct comparison of the number of science papers between observatories and even comparisons within a single observatory may not be the most effective or equitable way to analyze productivity. We do not advise assessing total publication output over time between facilities or even within a single observatory as the primary measure of science productivity. If such comparisons are required to satisfy oversight committees or funding agencies, observatory bibliography metrics should be provided with proper context. Caveats or limitations of the data and inherent differences in the datasets being compared should be provided along with the metrics report(s).
\item \textbf{Technology:} There are exciting technological advances in this field, including the use of machine learning to classify papers. However, it is important to emphasize that automation is never a substitute for human evaluation. Automated classification will need to be periodically inspected and validated by staff, and language models will need to be revised over time and then re-validated.
\item \textbf{Reports and Metrics:} Bibliographies are regularly included in reports to funding agencies and to observatory leadership. It is vital that there is clear documentation about how the bibliographies were compiled and what the metrics may or may not show.
\item \textbf{Staffing: }Many bibliographers work solo or as part of a small staff. Sharing of information, policies, and procedures through networking among bibliographers is vital to maintain the high standards, usability, and credibility of observatory bibliographies.
\item \textbf{New and Revised Bibliographies:} If a bibliography is not initiated at first light, it is difficult to begin one after years of observations without investing significant time to catch up. The examples of current observatory bibliographies shared throughout this paper and our guidelines and experiences can act as templates when devising or revising bibliographies.
\end{itemize}

\section{Discussion}
\label{sec:concl}
Observatory bibliographers play a vital role in demonstrating the scientific impact and value of an observatory. Our work is aligned with the goals and mission of our respective institutions, and consequently there is tremendous diversity in the ways we track publications and maintain publications databases. Our differences are driven by parameters such as observatory resources (personnel, time, money, IT capabilities), type of observatory, historical practices at an observatory, technical changes and evolution, and reporting requirements to outside agencies. Despite this diversity, there is a common set of best practices that guide our work. In this article, we provided examples and guidelines to construct and maintain robust and valuable observatory bibliographies. We also emphasized the importance of documenting, evaluating, continually refining, and sharing our bibliography practices with other bibliographers and the entire astronomical community. Changes implemented in external and internal systems, such as ADS and SciX or one's own data archive, may necessitate changes in observatory bibliography systems. As a result, bibliographers should expect to revise their metadata schema, processes, and other aspects of bibliographic work on a continual basis for the life of the bibliography.

\section*{Acknowledgments}
This research has made use of the Astrophysics Data System, funded by NASA under Cooperative Agreement 80NSSC21M00561. Raffaele D'Abrusco acknowledges support from the Chandra X-ray Center, which is operated by the Smithsonian Institution under NASA contract NAS8-03060. In addition to the ADS team, the authors wish to thank the anonymous reviewers for their helpful comments, insightful questions, and recommended edits.
\bibliography{OBC}

\begin{thebibliography}{}
\expandafter\ifx\csname natexlab\endcsname\relax\def\natexlab#1{#1}\fi
\providecommand{\url}[1]{\href{#1}{#1}}
\providecommand{\dodoi}[1]{doi:~\href{http://doi.org/#1}{\nolinkurl{#1}}}
\providecommand{\doeprint}[1]{\href{http://ascl.net/#1}{\nolinkurl{http://ascl.net/#1}}}
\providecommand{\doarXiv}[1]{\href{https://arxiv.org/abs/#1}{\nolinkurl{https://arxiv.org/abs/#1}}}

\bibitem[{{Blecksmith} {et~al.}(2005){Blecksmith}, {Bright}, {Rots}, {Winkelman}, {Green}, \& {Yukita}}]{2005ASPC..347..380B}
{Blecksmith}, S., {Bright}, J., {Rots}, A.~H., {et~al.} 2005, in Astronomical Society of the Pacific Conference Series, Vol. 347, Astronomical Data Analysis Software and Systems XIV, ed. P.~{Shopbell}, M.~{Britton}, \& R.~{Ebert}, 380.
\newblock \url{http://aspbooks.org/custom/publications/paper/347-0380.html}

\bibitem[{{Chen} {et~al.}(2022{\natexlab{a}}){Chen}, {Ebert}, {Mazzarella}, {Frayer}, {Terek}, {Chan}, {Cook}, {Lo}, {Schmitz}, \& {Wu}}]{2022PASP..134a4501C}
{Chen}, T.~X., {Ebert}, R., {Mazzarella}, J.~M., {et~al.} 2022{\natexlab{a}}, \pasp, 134, 014501, \dodoi{10.1088/1538-3873/ac3c36}

\bibitem[{{Chen} {et~al.}(2022{\natexlab{b}}){Chen}, {Schmitz}, {Mazzarella}, {Wu}, {van Eyken}, {Accomazzi}, {Akeson}, {Allen}, {Beaton}, {Berriman}, {Boyle}, {Brouty}, {Chan}, {Christiansen}, {Ciardi}, {Cook}, {D'Abrusco}, {Ebert}, {Frayer}, {Fulton}, {Gelino}, {Helou}, {Henderson}, {Howell}, {Kim}, {Landais}, {Lo}, {Loup}, {Madore}, {Monari}, {Muench}, {Oberto}, {Ocvirk}, {Peek}, {Perret}, {Pevunova}, {Ramirez}, {Rebull}, {Shemmer}, {Smale}, {Tam}, {Terek}, {Van Orsow}, {Vannier}, \& {Wang}}]{2022ApJS..260....5C}
{Chen}, T.~X., {Schmitz}, M., {Mazzarella}, J.~M., {et~al.} 2022{\natexlab{b}}, \apjs, 260, 5, \dodoi{10.3847/1538-4365/ac6268}

\bibitem[{{Chyla} {et~al.}(2015){Chyla}, {Accomazzi}, {Holachek}, {Grant}, {Elliott}, {Henneken}, {Thompson}, {Kurtz}, {Murray}, \& {Sudilovsky}}]{2015ASPC..495..401C}
{Chyla}, R., {Accomazzi}, A., {Holachek}, A., {et~al.} 2015, in Astronomical Society of the Pacific Conference Series, Vol. 495, Astronomical Data Analysis Software an Systems XXIV (ADASS XXIV), ed. A.~R. {Taylor} \& E.~{Rosolowsky}, 401, \dodoi{10.48550/arXiv.1503.05881}

\bibitem[{{Cutri} {et~al.}(2003){Cutri}, {Skrutskie}, {van Dyk}, {Beichman}, {Carpenter}, {Chester}, {Cambresy}, {Evans}, {Fowler}, {Gizis}, {Howard}, {Huchra}, {Jarrett}, {Kopan}, {Kirkpatrick}, {Light}, {Marsh}, {McCallon}, {Schneider}, {Stiening}, {Sykes}, {Weinberg}, {Wheaton}, {Wheelock}, \& {Zacarias}}]{2003tmc..book.....C}
{Cutri}, R.~M., {Skrutskie}, M.~F., {van Dyk}, S., {et~al.} 2003, {2MASS All Sky Catalog of Point Sources.} (NASA/IPAC Infrared Science Archive).
\newblock \url{https://irsa.ipac.caltech.edu/Missions/2mass.html}

\bibitem[{{Eichhorn}(2004)}]{2004A&G....45c...7E}
{Eichhorn}, G. 2004, A\&G, 45, 3.07, \dodoi{10.1046/j.1468-4004.2003.45307.x}

\bibitem[{{Grothkopf} {et~al.}(2018){Grothkopf}, {Meakins}, \& {Bordelon}}]{2018SPIE10704E..0RG}
{Grothkopf}, U., {Meakins}, S., \& {Bordelon}, D. 2018, in Society of Photo-Optical Instrumentation Engineers (SPIE) Conference Series, Vol. 10704, Observatory Operations: Strategies, Processes, and Systems VII, ed. A.~B. {Peck}, A.~L. {Seaman}, \& C.~R. {Benn}, 107040R, \dodoi{10.1117/12.2311667}

\bibitem[{{Henneken} \& {Kurtz}(2019)}]{2019AAS...23345301H}
{Henneken}, E., \& {Kurtz}, M.~J. 2019, in American Astronomical Society Meeting Abstracts, Vol. 233, American Astronomical Society Meeting Abstracts \#233, 453.01.
\newblock \url{https://baas.aas.org/abstracts}

\bibitem[{{Lagerstrom}(2015)}]{2015ASPC..492...99L}
{Lagerstrom}, J. 2015, in Astronomical Society of the Pacific Conference Series, Vol. 492, Open Science at the Frontiers of Librarianship, ed. A.~{Holl}, S.~{Lesteven}, D.~{Dietrich}, \& A.~{Gasperini}, 99.
\newblock \url{http://aspbooks.org/custom/publications/paper/492-0099.html}

\bibitem[{{Rots} {et~al.}(2012){Rots}, {Winkelman}, \& {Becker}}]{2012PASP..124..391R}
{Rots}, A.~H., {Winkelman}, S.~L., \& {Becker}, G.~E. 2012, \pasp, 124, 391, \dodoi{10.1086/665581}

\bibitem[{{Scire} {et~al.}(2010){Scire}, {Chan}, {Silbermann}, \& {Shields}}]{2010SPIE.7737E..1VS}
{Scire}, E., {Chan}, B. H.~P., {Silbermann}, N., \& {Shields}, A. 2010, in Society of Photo-Optical Instrumentation Engineers (SPIE) Conference Series, Vol. 7737, Observatory Operations: Strategies, Processes, and Systems III, ed. D.~R. {Silva}, A.~B. {Peck}, \& B.~T. {Soifer}, 77371V, \dodoi{10.1117/12.857735}

\bibitem[{{Scire} {et~al.}(2022){Scire}, {Rebull}, \& {Laine}}]{2022PASP..134e5001S}
{Scire}, E., {Rebull}, L., \& {Laine}, S. 2022, \pasp, 134, 055001, \dodoi{10.1088/1538-3873/ac4959}

\bibitem[{{Skrutskie} {et~al.}(2006){Skrutskie}, {Cutri}, {Stiening}, {Weinberg}, {Schneider}, {Carpenter}, {Beichman}, {Capps}, {Chester}, {Elias}, {Huchra}, {Liebert}, {Lonsdale}, {Monet}, {Price}, {Seitzer}, {Jarrett}, {Kirkpatrick}, {Gizis}, {Howard}, {Evans}, {Fowler}, {Fullmer}, {Hurt}, {Light}, {Kopan}, {Marsh}, {McCallon}, {Tam}, {Van Dyk}, \& {Wheelock}}]{2006AJ....131.1163S}
{Skrutskie}, M.~F., {Cutri}, R.~M., {Stiening}, R., {et~al.} 2006, \aj, 131, 1163, \dodoi{10.1086/498708}

\bibitem[{{van Raan}(2019)}]{10.1007/978-3-030-02511-3}
{van Raan}, A. 2019, Measuring Science: Basic Principles and Application of Advanced Bibliometrics, Springer Handbooks (Cham, Switzerland: Springer), 237--280, \dodoi{10.1007/978-3-030-02511-3}

\end{thebibliography}
\bibliographystyle{aasjournal}


\begin{appendix}
\section{Appendix A: Most Commonly Represented Journals for Astronomy Bibliographies} \label{sec:commonjournals}
\begin{center}
\begin{tabular}{ |c|c|c|c|c|} 
 \hline
 HST(1991--2021) & ESO(1996--present) & NRAO(1957--present) & Spitzer(2003--2020) Mid and far IR & NSF NOIRLab \\ 
 \hline
 ApJ 43\% & A\&A 46\% & ApJ 42.13\% & ApJ 43.4\% & MNRAS 35\% \\
 \hline
 MNRAS 19\% & MNRAS 23\% & MNRAS 17.68\% & MNRAS 19.5\% &ApJ 26\% \\
 \hline
 A\&A 14\% & ApJ 16\% & A\&A 17.67\% & A\&A 19.0\% &AJ 14\% \\
 \hline
 AJ 10\% & ApJL 5\% & AJ 8.62\% & AJ 6.4\% & A\&A 10\% \\
 \hline
 ApJS 2\% & AJ 4\% & ApJS 2.65\% & ApJS 4.1\% & ApJL 4\%\\
 \hline
 Icar 2\% & A\&ASuppl 1\% & Nature 1.6\% & PASJ 1.1\% & ApJS 4\% \\
 \hline
 PASP 1\% & ApJS 1\% & PASJ 0.94\% & PASP 0.8\% & PhRvD 1\% \\
 \hline
 Nature 1\% & Nature 1\% & Science 0.68\% & Nature and NatAs 1.1\% & PSJ 1\% \\
 \hline 
 & Icar 1\% & & Icar 0.6\% &\\
 \hline
 & AN 1\% & & AN, Ap\&SS, Science 0.2\% each &\\
 \hline
\end{tabular}
\end{center}
The share of Spitzer papers published in the ApJ dropped from 56.1\% in 2009 to 32.2\% in 2020, while the share published in MNRAS rose from 10.4\% to 28.0\%. This was due to changes in the funding profiles for observers granted time on the observatory. See \citet{2022PASP..134e5001S} for more information.

\section{Appendix B: Further Reading on Observatory Bibliographies} \label{sec:furthereading}
\begin{enumerate}
\item Abt, H. A. 2005, Information Obtainable from Bibliometric Studies, coas.conf, 2, \href{https://ui.adsabs.harvard.edu/abs/2005coas.conf....2A}{2005coas.conf....2A}

\item Accomazzi, A., Eichhorn, G. 2004, Publishing Links to Astronomical Data On-line, ASPC, 314, 181, \href{https://ui.adsabs.harvard.edu/abs/2004ASPC..314..181A}{2004ASPC..314..181A} 

\item Accomazzi, A., et al. 2012, Telescope Bibliographies: An Essential Component of Archival Data Management and Operations, SPIE, 8448, 84480K, \href{https://ui.adsabs.harvard.edu/abs/2012SPIE.8448E..0KA}{2012SPIE.8448E..0KA}

\item Alonso-Valdivielso, M. Á., Antonio, E. G. 2010, Why Include Bibliometric Analysis in the Activities of a Library Specialized in Astronomy? — Notes From the Libraries of INTA, ASPC, 433, 95, \href{https://ui.adsabs.harvard.edu/abs/2010ASPC..433...95A}{2010ASPC..433...95A} 

\item Bordelon, D., et al. 2016, Trends and Developments in VLT Data Papers as Seen Through telbib, SPIE, 9910, 99102B, \href{https://ui.adsabs.harvard.edu/abs/2016SPIE.9910E..2BB}{2016SPIE.9910E..2BB}

\item Cortes, R., Depoortere, D., Malaver, L. 2018, Astronomy in Chile: Assessment of Scientific Productivity Through a Bibliometric Analysis, EPJWC, 186, 05002, \href{https://ui.adsabs.harvard.edu/abs/2018EPJWC.18605002C}{2018EPJWC.18605002C} 

\item Crabtree, D. 2019, Canada's Astronomy Performance Based on Bibliometrics, clrp, 2020, 14, \href{https://ui.adsabs.harvard.edu/abs/2019clrp.2020...14C}{2019clrp.2020...14C} 

\item Erdmann, C., Grothkopf, U. 2010, Next Generation Bibliometrics and the Evolution of the ESO Telescope Bibliography, ASPC, 433, 81, \href{https://ui.adsabs.harvard.edu/abs/2010ASPC..433...81E}{2010ASPC..433...81E} 

\item Grothkopf, U., Meakins, S. 2012, ESO Telescope Bibliography: New Public Interface, Msngr, 147, 41, \href{https://ui.adsabs.harvard.edu/abs/2012Msngr.147...41G}{2012Msngr.147...41G} 

\item Grothkopf, U., Meakins, S. 2015, ESO telbib: Linking In and Reaching Out, ASPC, 492, 63, \href{https://ui.adsabs.harvard.edu/abs/2015ASPC..492...63G}{2015ASPC..492...63G} 

\item Grothkopf, U., et al. 2004, The ESO Telescope Bibliography on the Web: Linking Publications and Observations, AAS, 205, 182.06, \href{https://ui.adsabs.harvard.edu/abs/2004AAS...20518206G}{2004AAS...20518206G}

\item Grothkopf, U. 2011, Astronomy Libraries - Your Gateway to Information, EAS, 49, 107, \href{https://ui.adsabs.harvard.edu/abs/2011EAS....49..107G}{2011EAS....49..107G}

\item Grothkopf, U., Meakins, S. 2012, The ESO Telescope bibliography at Your Fingertips, SPIE, 8448, 844821, \href{https://ui.adsabs.harvard.edu/abs/2012SPIE.8448E..21G}{ 2012SPIE.8448E..21G}

\item Grothkopf, U., Novacescu, J. 2022, The Role of Astronomy Librarians in FAIR Bibliography Curation and Metric Analyses, astr.conf, 3, \href{https://ui.adsabs.harvard.edu/abs/2022astr.confE...3G}{2022astr.confE...3G}

\item Grothkopf, U., Meakins, S., Bordelon, D. 2015, If We Build It, Will They Come? Curation and Use of the ESO Telescope Bibliography, scop.conf, 1, \href{https://ui.adsabs.harvard.edu/abs/2015scop.confE..26G}{2015scop.confE..26G}

\item Grothkopf, U., Lagerstrom, J. 2011, Telescope Bibliometrics 101, ASSP, 1, 109, \href{https://ui.adsabs.harvard.edu/abs/2011ASSP...24..109G}{2011ASSP...24..109G}

\item Grothkopf, U., Meakins, S., Bordelon, D. 2018, Use Cases of the ESO Telescope Bibliography, EPJWC, 186, 06001, \href{https://ui.adsabs.harvard.edu/abs/2018EPJWC.18606001G}{2018EPJWC.18606001G}

\item Henneken, E. A., Kurtz, M. J. 2019, Usage Bibliometrics as a Tool to Measure Research Activity, hsti.book, 819, \href{https://ui.adsabs.harvard.edu/abs/2019hsti.book..819H}{2019hsti.book..819H}

\item Hourclé, J. 2014, Data Citation in Astronomy, lisa.conf, 21, \href{https://ui.adsabs.harvard.edu/abs/2014lisa.confP..21H}{2014lisa.confP..21H }

\item Kitt, S., Grothkopf, U. 2010, Telescope Bibliography Cookbook: Creating a Database of Scientific Papers That Use Observational Data, ASPC, 433, 109, \href{https://ui.adsabs.harvard.edu/abs/2010ASPC..433..109K}{2010ASPC..433..109K} 

\item Lagerstrom, J., et al. 2012, Observatory Bibliographies: Current Practices, SPIE, 8448, 84481X, \href{https://ui.adsabs.harvard.edu/abs/2012SPIE.8448E..1XL}{2012SPIE.8448E..1XL}

\item Madrid, J. P., Macchetto, F. D. 2007, A Method to Measure the Scientific Output of the Hubble Space Telescope, ASPC, 377, 79, \href{https://ui.adsabs.harvard.edu/abs/2007ASPC..377...79M}{2007ASPC..377...79M} 

\item Meakins, S., Grothkopf, U. 2012, Linking Publications and Observations: The ESO Telescope Bibliography, ASPC, 461, 767, \href{https://ui.adsabs.harvard.edu/abs/2012ASPC..461..767M}{2012ASPC..461..767M }

\item Meakins, S., et al. 2014, Two Uears of ALMA bibliography: Lessons Learned, SPIE, 9149, 914926, \href{https://ui.adsabs.harvard.edu/abs/2014SPIE.9149E..26M}{2014SPIE.9149E..26M} 

\item National Research Council. 2012. For Attribution: Developing Data Attribution and Citation
Practices and Standards: Summary of an International Workshop. Washington, DC: The National
Academies Press \href{https://doi.org/10.17226/13564}{DOI:10.17226/13564}

\item Perret, E., et al. 2018, Shared Nomenclature and Identifiers for Telescopes and Instruments, EPJWC, 186, 04002, \href{https://ui.adsabs.harvard.edu/abs/2018EPJWC.18604002P}{2018EPJWC.18604002P}

\item Rots, A. H., Winkelman, S. L., Becker, G. E. 2012, Meaningful Metrics for Observatory Publication Statistics, SPIE, 8448, 84480J, \href{https://ui.adsabs.harvard.edu/abs/2012SPIE.8448E..0JR}{2012SPIE.8448E..0JR}

\item Savaglio, S., Grothkopf, U. 2014, Swift Publication Statistics and the Comparison with Other Major Observatories, lisa.conf, 4, \href{https://ui.adsabs.harvard.edu/abs/2014lisa.confP...4S}{2014lisa.confP...4S}

\item Schmitz, M., et al. 1995, A Uniform Bibliographic Code, VA, 39, 272, \href{https://ui.adsabs.harvard.edu/abs/1995VA.....39R.272S}{1995VA.....39R.272S}

\item Sterzik, M., et al. 2016, The Impact of Science Operations on Science Return at the Very Large Telescope, SPIE, 9910, 991003, \href{https://ui.adsabs.harvard.edu/abs/2016SPIE.9910E..03S}{2016SPIE.9910E..03S} 

\item Stevens-Rayburn, S., Grothkopf, U. 2007, Creating Telescope Bibliographies Electronically -- Are We There Yet?, ASPC, 377, 53, \href{https://ui.adsabs.harvard.edu/abs/2007ASPC..377...53S}{2007ASPC..377...53S} 

\item Winkelman, S., Rots, A. 2012, Observatory Bibliographies: Not Just for Statistics Anymore, SPIE, 8448, 844829, \href{https://ui.adsabs.harvard.edu/abs/2012SPIE.8448E..29W}{2012SPIE.8448E..29W}

\item Wynholds, L. 2011, Linking to Scientific Data: Identity Problems of Unruly and Poorly Bounded Digital Objects, International Journal of Digital Curation, 6, 214  \href{https://doi.org/10.2218/ijdc.v6i1.183}{DOI:10.2218/ijdc.v6i1.183} 
\end{enumerate}

\section{Appendix C: Abbreviations \& Acronyms} \label{sec:abbreviations}

\noindent{\textbf{2MASS}: Two Micron All Sky Survey}\\
\textbf{ADS}: Astrophysics Data System\\
\textbf{AI}: artificial intelligence\\
\textbf{AIP}: American Institute of Physics\\
\textbf{AJ}:  \textit{Astronomical Journal}\\
\textbf{Ap\&SS}:  \textit{Astrophysics and Space Sciences}\\
\textbf{API}: application program interface\\
\textbf{ApJ}: \textit{The Astrophysical Journal}\\
\textbf{ApJL}: \textit{The Astrophysical Journal Letters}\\
\textbf{ApJS}: \textit{The Astrophysical Journal Supplement}\\
\textbf{ASP}: Astronomical Society of the Pacific\\
\textbf{ATST}: Advanced Technology Solar Telescope\\
\textbf{AUI}: Associated Universities Inc.\\
\textbf{AURA}: Association of Universities for Research Astronomy, Inc.\\
\textbf{AXAF}: Advanced X-ray Astrophysics Facility, previous name of Chandra\\
\textbf{bibcode}: bibliographic code [ADS]\\
\textbf{bibgroup}: bibliographic group [ADS]\\
\textbf{CfA}: Center for Astrophysics | Harvard \& Smithsonian\\
\textbf{CLASH}: HST Legacy Program\\
\textbf{COS}: Cosmic Origins Spectrograph [HST]\\
\textbf{COSMOS}: Cosmological Evolution Survey; HST Legacy Program\\
\textbf{CSP}: Chandra Science Paper\\
\textbf{CTIO}: Cerro Tololo Inter-American Observatory\\
\textbf{CXC}: Chandra X-ray Center\\
\textbf{DKIST}: Daniel K. Inouye Solar Telescope \\
\textbf{DMSO}: Data Science Mission Office [STScI]\\
\textbf{DOI}: digital object identifier\\
\textbf{ELT}: Extremely Large Telescope\\
\textbf{ESO}: European Southern Observatory\\
\textbf{FTE}: full-time equivalent\\
\textbf{GALEX}: Galaxy Evolution Explorer\\
\textbf{GLASS}: HST Legacy Program\\
\textbf{GOALS}: Legacy Spitzer Space Telescope program\\
\textbf{GTO}: Guaranteed Time Observations\\
\textbf{GUI}: Graphical User Interface\\
\textbf{HST}: Hubble Space Telescope\\
\textbf{IAC}: Instituto de Astrofísica de Canarias\\
\textbf{IAU}: International Astronomical Union\\
\textbf{ID}: identifier\\
\textbf{IRAC}: Infrared Array Camera; Spitzer Space Telescope instrument\\
\textbf{IRAS}: Infrared Astronomical Satellite\\
\textbf{IRS}: Infrared Spectrograph; Spitzer Space Telescope instrument\\
\textbf{IRSA}: Infrared Science Archive\\
\textbf{IT}: Information technology\\
\textbf{IUE}: International Ultraviolet Explorer\\
\textbf{JSON}: JavaScript Object Notation\\
\textbf{JWST}: James Webb Space Telescope, formerly Next Generation Space Telescope (NGST)\\
\textbf{MAST}: Mikulski Archive for Space Telescopes [HST]\\
\textbf{MIPS}: Multiband Imaging Photometer; Spitzer Space Telescope instrument\\
\textbf{ML}: machine learning\\
\textbf{MNRAS}: \textit{Monthly Notices of the Royal Astrophysical Society}\\
\textbf{NAOJ}: National Astronomical Institute of Japan\\
\textbf{NASA}: National Aeronautics and Space Administration\\
\textbf{NatAs}: \textit{Nature Astronomy}\\
\textbf{NED}: NASA/IPAC Extragalactic Database\\
\textbf{NEOWISE}: NASA mission\\
\textbf{NGRST}: Nancy Grace Roman Space Telescope, formerly Wide Field Infrared Telescope (WFIRST)\\
\textbf{NGST}: Next Generation Space Telescope, former name of the James Webb Space Telescope (JWST)\\
\textbf{NIRSpec}: Near Infrared Spectrograph [JWST]\\
\textbf{NOAO}: National Optical Astronomy Observatory\\
\textbf{NOIRLab}: U.S. National Science Foundation National Optical-Infrared Astronomy Research Laboratory\\
\textbf{NRAO}: National Radio Astronomy Observatory\\
\textbf{OBC}: Observatory Bibliographers Collaboration \\
\textbf{PASJ}: \textit{Publications of the Astronomical Society of Japan}\\
\textbf{PASP}: \textit{Publications of the Astronomical Society of the Pacific}\\
\textbf{PI}: principal investigator\\
\textbf{postdoc}: postdoctoral researcher\\
\textbf{RTS}: Roman Space Telescope, abbreviated acronym for the Nancy Grace Roman Space Telescope, formerly the Wide Field Infrared Survey Telescope (WFIRST)\\
\textbf{SAO}: Smithsonian Astrophysical Observatory\\
\textbf{SciX}: NASA Science Explorer\\
\textbf{SIMPLE}: Spitzer IRAC/MUSYC Public Legacy Survey in the Extended Chandra Deep Field South; legacy Spitzer Space Telescope program\\
\textbf{SMD}: Science Mission Directorate\\
\textbf{SPIE}: Society of Photographic Instrumentation Engineers\\
\textbf{STScI}: Space Telescope Science Institute\\
\textbf{TESS}: Transiting Exoplanet Survey Satellite\\
\textbf{UI}: user interface\\
\textbf{VLT}: Very Large Telescope [ESO]\\
\textbf{WFIRST}: Wide Field Infrared Survey Telescope (former name of Nancy Grace Roman Space Telescope, NGRST or RST)\\
\textbf{WISE}: Wide-field Infrared Survey Explorer\\
\end{appendix}

\end{document}